\documentstyle[12pt] {article}
\begin {document}
\parindent=15pt
\begin{center}
\vskip 1.5 truecm
{\bf STRIPPING CROSS SECTIONS IN COLLISIONS OF LIGHT NUCLEI}\\
\vspace{.5cm}
Yu.M.Shabelski \\
\vspace{.5cm}
Petersburg Nuclear Physics Institute,\\
Gatchina, St.Petersburg 188350 Russia \\
\vspace{.5cm}
and \\
\vspace{.5cm}
D.Treleani \\
\vspace{.5cm}
Dipartimento di Fisica Teorica dell'Universit\`a and INFN,\\
Trieste, I 34014 Italy\\
\end{center}
\vspace{1cm}
\begin{abstract}
By evaluating all the contributions of the intermediate
states of the
multiple scattering theory diagrams, we compute the integrated stripping
cross sections of collisions among light nuclei. The resulting
expressions have the simple form of a combination of total inelastic
cross sections of nuclear reactions with projectile nuclei differing in
the atomic mass number. We also check the accuracy of some widely used
relations in heavy ion collisions.

\end{abstract}
\vspace{3cm}

E-mail shabel@vxdesy.desy.de \\

E-mail daniel@trieste.infn.it \\


\newpage

\section{Introduction}

Some exclusive cross sections can be expressed as differences of two (or
more) total inelastic cross sections. In ref.\cite{ASS} the probability
for one constituent quark of a proton to interact in a collision with
nucleus A, while the other two constituent quarks are spectators, was
written, in a simplest probabilistic approach, as
\begin{equation}
V_1^p(A) = \frac{3}{\sigma_{prod}^{pA}} \int d^2b
e^{-2\sigma_q T(b)} [1 - e^{-\sigma_q T(b)}] \;,
\end{equation}
where $\sigma_q \approx \frac{1}{3} \sigma^{pp}_{inel} \approx 10$ mb,
\begin{equation}
T(b) = A \int_{-\infty}^{+\infty} dz \rho (b,z)
\end{equation}
and $\rho (r = \sqrt{b^2 + z^2})$ is nuclear density.

Eq. (1) can be obviously rewritten as
\begin{equation}
V_1^p(A) = \frac{3}{\sigma_{prod}^{pA}}
\Bigl[\int d^2b (1 - e^{-3\sigma_q T(b)}) -
\int d^2b (1 - e^{-2\sigma_q T(b)})
\Bigr] \;,
\end{equation}
where the two integrals represent the absorption cross sections of a
three-quark and of a two-quark system respectively.

In the present note we show that in the case of collisions among two
light nuclei similar expressions hold to a very high accuracy,
accounting in fact for the contributions of interference terms, loop
diagrams, etc. The processes which we consider are nuclear
reactions with stripping of one or few nucleons
and we compute
the corresponding integrated cross
sections and the average number of interacting nucleons.
The main tools of our analysis are the multiple
scattering theory (Glauber theory) and the AGK cutting rules \cite{AGK}.
We study first the case of tritium-deuteron ($td$) interaction and
we discuss afterwards how the results are generalized to the case of
heavier nuclei.

The region of validity of the present analysis is the
energy region where the multiple scattering
theory is available, namely from a few hundred MeV per nucleon
to superhigh energies, where
inelastic screening corrections have to be taken into account.
For simplicity we neglect the difference
between $pp$ and $pn$ interactions.

\section{Multiple scattering theory diagrams for the total cross section}

The simplest nuclear target is deuteron. One can therefore start
considering the elastic scattering amplitude, or the total cross
section, for $pd$, $dd$ and $td$ interactions. The multiple scattering
theory diagrams for the processes are shown in Figs. 1, 2 and 3, where
the interacting nucleons are represented by dots and the interactions by
solid lines. Non-interacting nucleons are not shown. The number at a
side of each diagram is the corresponding combinatorial factor $l_i$ and
$\Delta_{n,i}$ is the contribution of the diagram to the total cross
section. The index $n$ in $\Delta_{n,i}$ represents the number of
interactions and the index $i$ allows one to distinguish between
different diagrams at the same order in the number of interactions. As
an example, in Fig. 3 the combinatorial factor $l_i$ is equal to 6 for
the diagram corresponding to $\Delta_{2,b}$, since there are three
possibilities to choose the non-interacting nucleon in tritium and two
possibilities to choose the interacting nucleon in deuteron.

The total cross section of two interacting nuclei is represented by the
sum of all possible diagrams in the multiple scattering theory, which,
by introducing the quantities
\begin{equation}
\Delta_n = \sum_i l_i \Delta_{n,i} \;,
\end{equation}
is expressed as:
\begin{equation}
\sigma_{tot} = \sum_n^{A \cdot B} (-1)^{n-1} \Delta_n \;.
\end{equation}

\section{ Cross section for stripping one nucleon in $td$ collisions}

The simplest stripping process to consider is the one where one nucleon,
of a fast $t$-nucleus, is
scattered with a large transverse momentum with respect to its Fermi
momentum, while the two
other nucleons are spectators. From the point of view of the unitarity
relation the needed intermediate states can be found when the
nucleon-nucleon interaction blob, Fig. 4a, is cut. In the case of the
cut between two interactions, as in Fig. 4b, the probability to find a
nucleon with comparatively large transverse momentum is suppressed
by the nuclear form factor.

The cross section corresponding to a given cut diagram is deternimed
by the AGK cutting rules \cite{AGK,Sh1,Sh2} which were first obtained for
reggeon diagrams \cite{AGK} and later were proved for
hadron-nucleus \cite{Sh1} and nucleus-nucleus \cite{Sh2} interactions.
As an example, the cross section of the process corresponding to a cut
diagram of the kind shown in Fig. 4a is equal to the contribution to the
total cross section of the same uncut diagram ($\Delta_{2c}$ in the
considered case) multiplied by the AGK \cite{AGK,Sh1,Sh2} factor
\begin{equation}
D_n^{(k)} = (-1)^{n+k} 2^{n-1} \frac{n!}{(n-k)! k!} \;,
\end{equation}
where $n$ is the number of interactions and $k$ the number of cut
blobs. In Fig. 4a $n = 2$ and $k = 1$.

The one-cut-blob contribution to the cross section for stripping
one-nucleon in tritium-deuteron interactions, in accordance with the
diagrams in Fig. 3 and with Eq. (6), is expressed as
\begin{eqnarray}
\sigma_1^{(1)} & = & 6 \Delta_1 - 4 (3 \Delta_{2,a} + 6 \Delta_{2,b} +
6 \Delta_{2,c}) +12 (12 \Delta_{3,a} + 2 \Delta_{3,b} + 6 \Delta_{3,c})
- \\ \nonumber
& - & 32 (3 \Delta_{4,a} + 6 \Delta_{4,b} + 6 \Delta_{4,c})
+ 480 \Delta_5 - 192 \Delta_6 \;.
\end{eqnarray}

A different way to produce a stripped nucleon is to cut
two blobs as in Fig. 3 (2a). A further possibility is to cut two blobs
in Fig. 3 (3a). In this last case one needs however to include an
additional factor 1/3 since only one combination out of the possible
three gives the needed state.  After accounting for all factors, the
contribution of the two-cut-blob terms to the one-nucleon stripping
cross section in tritium-deuteron reactions is written as
\begin{equation}
\sigma_1^{(2)} = 6 \Delta_{2,a} - 48 \Delta_{3,a} + 48 (\Delta_{4,a} +
\Delta_{4,b} + \Delta_{4,c}) - 192 \Delta_5 + 96 \Delta_6 \;.
\end{equation}

All other possible cuts of the multiple scattering theory diagrams
give more than one stripped nucleon.
The total cross section for one-nucleon-stripping reactions, $\sigma_1$,
is therefore equal to the sum $\sigma_1^{(1)} + \sigma_1^{(2)}$ which
can be written as
\begin{eqnarray}
\sigma_1 & = & 3 [2 \Delta_1 - (2
\Delta_{2,a} + 8 \Delta_{2,b} + 8 \Delta_{2,c}) + (32 \Delta_{3,a} + 8
\Delta_{3,b} + 24 \Delta_{3,c}) - \\ \nonumber & - & 8 (2 \Delta_{4,a} +
6 \Delta_{4,b} + 6 \Delta_{4,c}) + 96 \Delta_5 - 32 \Delta_6] \;.
\end{eqnarray}

The total inelastic $td$ and $dd$ cross sections
can be expressed in the form \cite{Sh2} (also see below) :
\begin{equation}
\sigma_{inel} = \sum_n^{A \cdot B} (-1)^{n-1} 2^{n-1} \Delta_n
\end{equation}
The values of $\Delta_n$ are different in $td$ and in $dd$ interactions.
One may however notice
that the value of each $\Delta_{n,i}$, in $td$ interactions, is
determined by the nucleon-nucleon cross section parameters and by the
radii of tritium and deuteron, which are present in the result with
coefficients depending on the kind of diagram considered. If one forces
one of the terms $\Delta_{n,i}$ to be the same in $td$ and in $dd$
interactions, by rescaling for example the deuteron radius, all other
terms $\Delta_{n,i}$, when rescaled in the same way, are therefore also
forced to coincide. One can then compute the $dd$ process in multiple
scattering theory by changing the radius of the projectile deuteron in
such a way as to obtain equal values for all $\Delta_{n,i}$ in $td$ and
in $dd$ reactions. Obviously one cannot compare the value of
$\sigma_{inel}^{dd}$ which is obtained in this way with the experimental
value. The expression of $\sigma_{inel}^{dd}$, that one obtains with
this procedure, allows however one to write in a rather compact form
the stripping cross section in terms of the difference
$\sigma_{inel}^{td} - \sigma_{inel}^{dd}$. As a consequence of the
cancellation between many of the terms appearing in $\sigma_{inel}^{td}$
and in $\sigma_{inel}^{dd}$ one can in fact write
\begin{eqnarray}
\sigma_{inel}^{td} - \sigma_{inel}^{dd} & = & 2 \Delta_1 - (2 \Delta_{2,a}
+ 8 \Delta_{2,b} + 8 \Delta_{2,c}) + (32 \Delta_{3,a} + 8 \Delta_{3,b} +
\\ \nonumber
& + & 24 \Delta_{3,c}) - 8 (2 \Delta_{4,a} + 6 \Delta_{4,b} +
6 \Delta_{4,c}) + 96 \Delta_5 - 32 \Delta_6 \;
\end{eqnarray}
which is precisely one third of $\sigma_1$ as expressed in Eq.(9):
\begin{equation}
\sigma_1 = 3 (\sigma_{inel}^{td} - \sigma_{inel}^{dd}) \;.
\end{equation}

\section{Cross section for stripping two nucleons in $td$ collisions}

The cross section $\sigma_2$, for stripping two nucleons
of tritium, in $td$ interaction, can be computed analogously to
the computation of $\sigma_1$
in the previous paragraph. In
the process the $t$-nucleus
dissociates into one spectator nucleon and two nucleons with
transverse momentum large as compared with their Fermi momentum.
The cuts to be taken into account in the multiple scattering theory
diagrams are those involving two tritium
nucleons. One obtains:
\begin{eqnarray}
\sigma_2^{(2)} & = & 2 (6 \Delta_{2,b} + 6 \Delta_{2,c}) -
12 (8 \Delta_{3,a} + 2 \Delta_{3,b} + 6 \Delta_{3,c}) + \\ \nonumber
& + & 48 (2 \Delta_{4,a} + 5 \Delta_{4,b} + 5 \Delta_{4,c})
- 768 \Delta_5 + 384 \Delta_6 \;.
\end{eqnarray}

Analogously to the case discussed in the previous Sect., one has to keep
into account also of the contribution where the two stripped nucleons
are originated in processes where three or more interaction blobs are
cut. The simplest example is the diagram in Fig. 3 (3a). The
contribution to $\sigma_2$ from three cut blobs is equal to
\begin{equation}
\sigma_2^{(3)} = 48 \Delta_{3,a} - 32 (3 \Delta_{4,a} + 3 \Delta_{4,b}
+ 3 \Delta_{4,c}) + 576 \Delta_5 - 384 \Delta_6 \;.
\end{equation}
In a similar way the contribution from four cut blobs is :
\begin{equation}
\sigma_2^{(4)} = 24 \Delta_{4,a} - 96 \Delta_5 + 96 \Delta_6 \;.
\end{equation}

The expression of the sum of all contributions, where two tritium
nucleons are stripped in $td$ interaction,
$\sigma_2 = \sum \sigma_2^{(i)}$, is therefore
\begin{eqnarray}
\sigma_2 & = & 3 [(4 \Delta_{2,b} + 4 \Delta_{2,c}) -
(16 \Delta_{3,a} + 8 \Delta_{3,b} + 24 \Delta_{3,c}) + \\ \nonumber
& + & (8 \Delta_{4,a} + 48 \Delta_{4,b} + 48 \Delta_{4,c})
- 96 \Delta_5 + 32 \Delta_6] \;.
\end{eqnarray}
As for $\sigma_1$ in the previous paragraph,
by making use of Eq. (10) one obtains
\begin{equation}
\sigma_2 = 3 (2\sigma_{inel}^{dd} - \sigma_{inel}^{td} -
\sigma_{inel}^{pd}) \;.
\end{equation}

\section{ Cross section for stripping three nucleons in $td$ collisions }

The cross section of the process where all the three tritium-nucleons
are scattered with a transverse momentum larger with respect to their
Fermi momentum, $\sigma_3$, can be calculated analogously :
\begin{equation}
\sigma_3^{(3)} = 4 (2 \Delta_{3,b} + 6 \Delta_{3,c})
- 32 (3 \Delta_{4,b} + 3 \Delta_{4,c}) + 384 \Delta_5 - 256 \Delta_6 \;,
\end{equation}
\begin{equation}
\sigma_3^{(4)} = 8 (6 \Delta_{4,b} + 6 \Delta_{4,c}) - 384 \Delta_5
+ 384 \Delta_6 \;,
\end{equation}
\begin{equation}
\sigma_3^{(5)} = 96 \Delta_5 - 192 \Delta_6 \;,
\end{equation}
\begin{equation}
\sigma_3^{(6)} = 32 \Delta_6 \;.
\end{equation}
and
\begin{equation}
\sigma_3 = \sum \sigma_2^{(i)} = (8 \Delta_{3,b} + 24 \Delta_{3,c}) -
(48 \Delta_{4,b} + 48 \Delta_{4,c}) + 96 \Delta_5 - 32 \Delta_6 \;
\end{equation}
which can be written in the form
\begin{equation}
\sigma_3 = \sigma_{inel}^{td} - 3 \sigma_{inel}^{dd} +
3 \sigma_{inel}^{pd} \;.
\end{equation}

\section{Test of a few relations in nucleus-nucleus interactions}

The equations derived in the previous paragraphs are of rather general
validity in multiple scattering theory. One has therefore the
possibility to test a few well known relations which are widely
used in nucleus-nucleus interactions.

The total inelastic tritium-deuteron cross
section is equal to the sum of
$\sigma_1$, $\sigma_2$ and $\sigma_3$. By using the explicit expressions
in Eq.s (9), (16) and  (22) one can write:
\begin{eqnarray}
\sigma_{inel}^{td} & = & 6 \Delta_1 - (6 \Delta_{2,a} + 12 \Delta_{2,b} +
12 \Delta_{2,c}) + (48 \Delta_{3,a} + 8 \Delta_{3,b} + \\ \nonumber
& + & 24 \Delta_{3,c}) -(24 \Delta_{4,a} + 48 \Delta_{4,b} +
48 \Delta_{4,c}) + 96 \Delta_5 - 32 \Delta_6 \;,
\end{eqnarray}
which, recalling the definition of $\Delta_n$ in Eq.(4), shows that
Eq. (10) is an exact relation in multiple scattering theory.

A similar conclusion holds for the average number of interacting
nucleons of the incident tritium nucleus, $<\!N_A\!>$. To obtain
the average number one in fact writes
\begin{equation}
\sigma_{inel}^{td} <\!N\!>  = \sigma_1 + 2 \sigma_2 + 3 \sigma_3
\end{equation}
and, after substituting the explicit expressions of $\sigma_1$,
$\sigma_2$ and $\sigma_3$, one obtains
\begin{equation}
\sigma_1 + 2 \sigma_2 + 3 \sigma_3 =
6 \Delta_1 - 6 \Delta_{2,a} = 3 \sigma_{inel}^{pd}
\end{equation}
in agreement with the general expression of $<\!N_A\!>$ \cite{BBC}.

\section{More general expressions}

We have performed analogous calculations in the case of interactions
among various different pairs of light nuclei
($dd, d\,^4\!He, tt, t\,^4\!He$). In all cases the results are the same,
namely they are of the kind of Eqs. (12), (17) and (23). The cross
sections for stripping one, two, three, etc. nucleons from nucleus $A$,
in collisions among the two nuclei $A$ and $B$, are written as
\begin{equation}
\sigma_1^{AB} = A (\sigma_{inel}^{A,B} - \sigma_{inel}^{A-1,B}) \;,
\end{equation}
\begin{equation}
\sigma_2^{AB} = \frac{A(A-1)}{2!} (-\sigma_{inel}^{A,B} +
2 \sigma_{inel}^{A-1,B} - \sigma_{inel}^{A-2,B}) \;,
\end{equation}
\begin{equation}
\sigma_3^{AB} = \frac{A(A-1)(A-2)}{3!} (\sigma_{inel}^{A,B} -
3 \sigma_{inel}^{A-1,B} + 3 \sigma_{inel}^{A-2,B} -
\sigma_{inel}^{A-3,B})
\end{equation}
with an obvious generalization to the case where the number of stripped
nucleons is larger than three. As in the case of tritium, previously
discussed, in the expressions above the inelastic cross sections have to
be computed by taking, for the nuclei with atomic mass number $A-1$,
$A-2$ and $A-3$, the same radius as for the nucleus with atomic mass
number $A$. 

We have also checked that the expression giving the average number of
interacting nucleons of the projectile nucleus $A$, actually \cite{BBC}
\begin{equation}
<\!N\!>_{AB}  = A \sigma_{inel}^{NB}/ \sigma_{inel}^{AB},
\end{equation}
is exact at all orders in multiple scattering theory.

\section{Conclusions}

By making use of the AGK cutting rules we have computed the stripping
cross sections in collisions involving light nuclei. The expressions
which we have obtained can be formally written in a compact way as
differences between total inelastic cross sections of reactions where
the projectile nuclei differ in the atomic mass number. The total
inelastic cross sections, which are used to express the stripping cross
sections, are not the experimental quantities. They represent formal
expressions to be evaluated in multiple scattering theory, by using
the same value of nuclear radius for the different projectile nuclei.
All stripping cross sections, in interactions among light nuclei, are
therefore written in a convenient way as combinations of the same
quantities, which are easily computed in the multiple scattering
theory formalism. Multiple scattering theory and the AGK cutting
rules allow therefore one fix the normalization of the spectra of the
stripped nucleons. The normalization in fact is not a minor problem
\cite{BT}, because of the difficulty of constructing explicitly
the ortonormal set of scattering states of a process where various
different scattering channels and bound states are present. Which is
precisely the case of the stripping reactions.  One in fact obtains a
non-trivial link between the different stripping cross sections, namely
the relations (27), (28) and (29).

In the present note we have also tested some widely used relations in
nuclear collisions, Eq.s (10) and (29), which have been shown to be
accurate at all orders in the multiple scattering theory.

Some numerically small corrections to all
our results can appear from the contributions of cuts between
nucleon-nucleon interaction blobs, as shown in Fig. 4b, see discussion
in Sect. 3. One has also to point out that, as one
can see from Eq. (24), the numerically small
cross section of "stripping zero nucleon", i.e. of the process
$t + d \to t + p + n$, is not included into $\sigma^{td}_{inel}$.

We finally remark that, while the expressions which we have worked out
refer to the stripping cross sections, our approach can be easily
generizable to reactions of different kind, as for example pion
production and Drell-Yan pair production.

\vskip.1in
{\bf Acknowledgements}
\vskip.1in

This work is supported in part by INTAS grant 93-0079 and by the Italian
Ministry of University and of Scientific and Technological Research, by
means of the Fondi per la Ricerca scientifica - Universit\`a di Trieste.

\vspace{1cm}

\begin{center}
{\bf Figure captions}\\
\end{center}

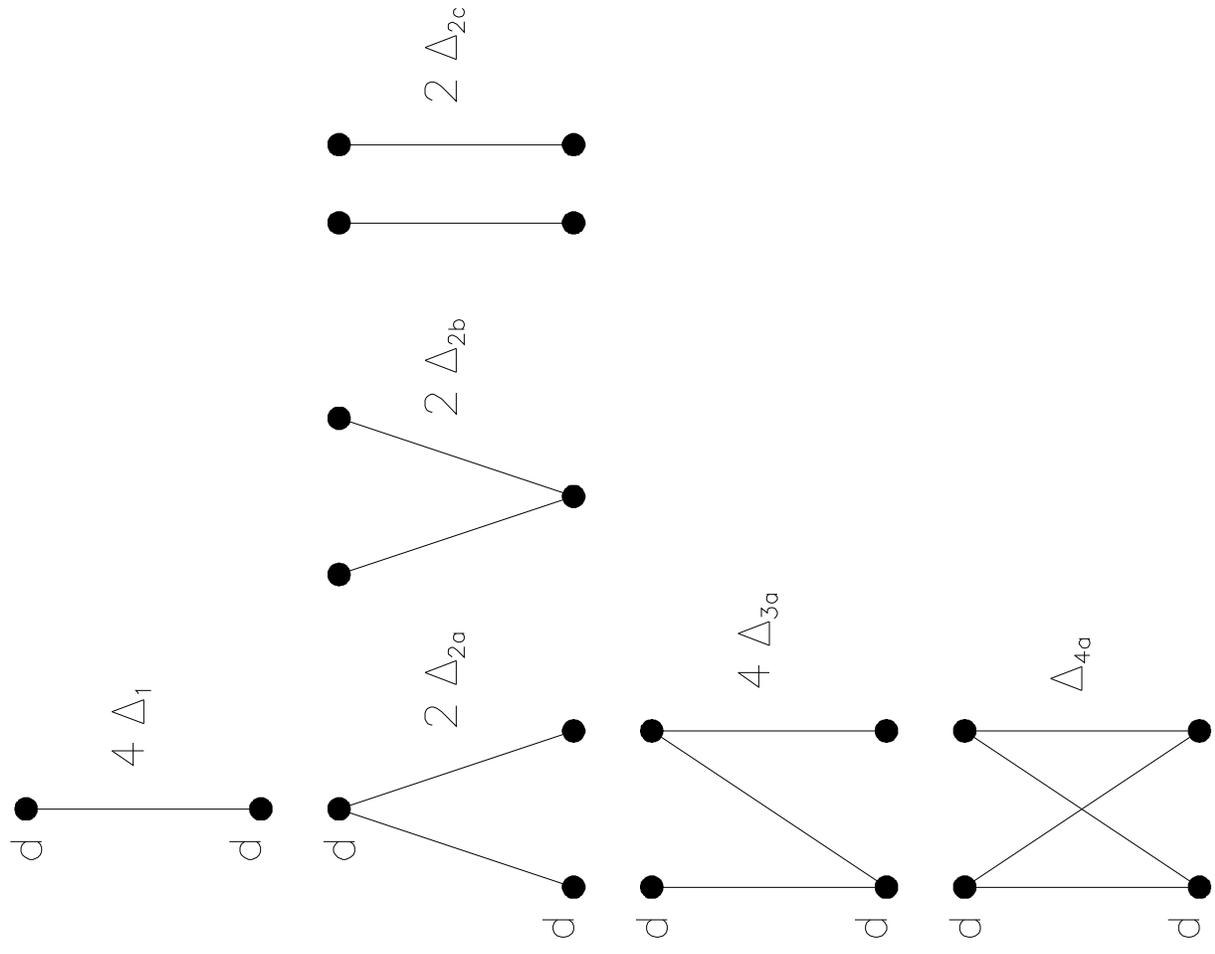
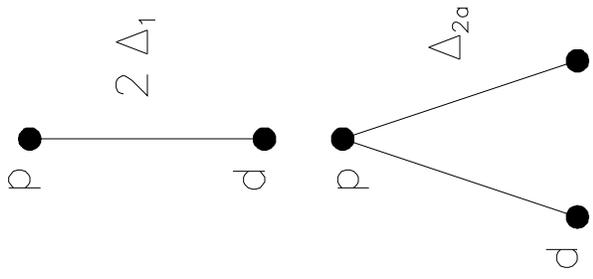
Fig. 1. Multiple scattering theory diagrams for $pd$ elastic scattering
amplitude. (See text for definitions.)

Fig. 2. Multiple scattering theory diagrams for $dd$ elastic scattering
amplitude. (See text for definitions.)

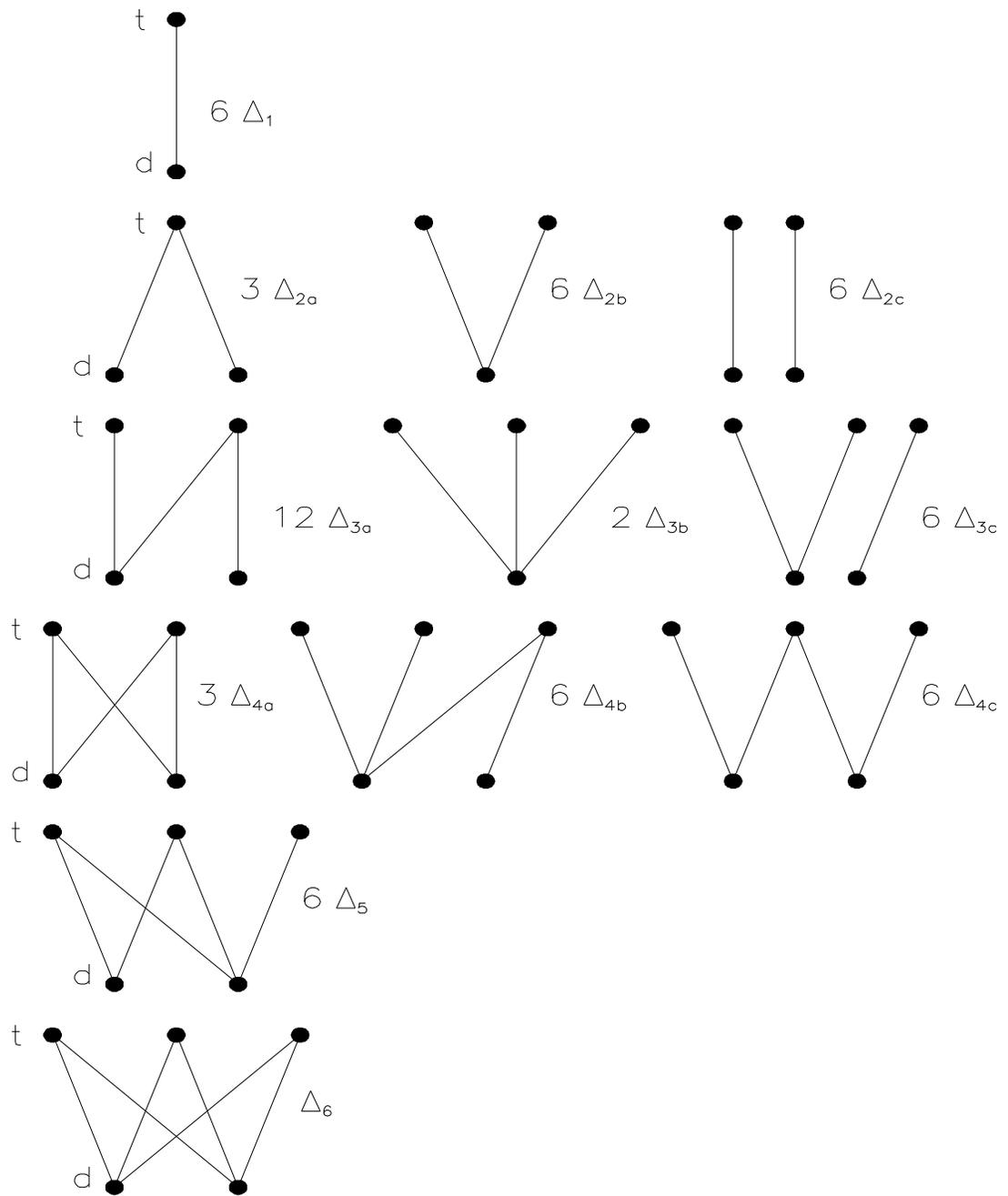
Fig. 3. Multiple scattering theory diagrams for $td$ elastic scattering
amplitude. (See text for definitions.)

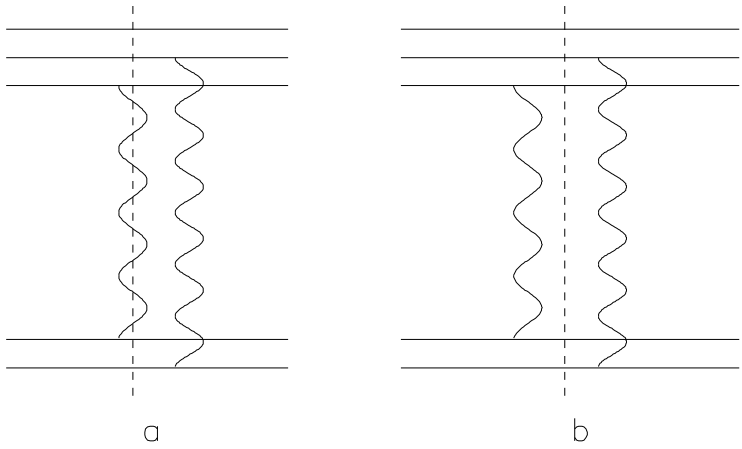
Fig. 4. The examples of cut of the diagram Fig. 3 (2c): cut of the
blob of nucleon-nucleon interaction (a) and cut between blobs (b).


\end{document}